\documentclass[
preprint,
superscriptaddress,
nofootinbib,
natbib,
amsmath,amssymb,
aps,
pra,
]{revtex4-1}

\usepackage{graphicx}
\usepackage{float} 
\usepackage[caption=false,font=footnotesize]{subfig}  
\usepackage{dcolumn}
\usepackage{bm}
\usepackage{verbatim}   
\usepackage{color}      
\usepackage{siunitx} 
\usepackage[capitalise]{cleveref} 
\usepackage{verbatim} 

\begin{document}

\title{Third-order response of metallic acGNR to an elliptically-polarized
terahertz excitation field}

\author{Yichao Wang}
\affiliation{Electrical and Computer Engineering}
\author{David R. Andersen}
\affiliation{Electrical and Computer Engineering}
\affiliation{Physics and Astronomy\\The University of Iowa, Iowa City, IA 52242 USA}

\date{Jun. 15, 2016}

\begin{abstract}
We present a theoretical description of the third-order response
induced by an elliptically-polarized terahertz
beam normally-incident on intrinsic and extrinsic metallic armchair graphene
nanoribbons.
Our results show that using a straightforward experimental setup, it should be
possible to observe novel polarization-dependent nonlinearities at low
excitation field strengths of the order of $10^4 \, \mathrm{V/m}$. At
low temperatures the Kerr nonlinearities in extrinsic nanoribbons
persist to significantly higher
excitation frequencies than they do for linear polarizations, and at room
temperatures, the third-harmonic nonlinearities are enhanced by 2-3 orders of
magnitude.
Finally, the Fermi-level and temperature dependence of the nonlinear response is
characterized.
\end{abstract}
\maketitle
Graphene is a flat monolayer of carbon atoms tightly packed into a 2D honeycomb lattice. Graphene has emerged to be a very promising candidate for terahertz (\si{THz}) applications, and opens up the possibility of graphene based devices for \si{THz} optoelectronic and
photonic applications \cite{review2014}. Theoretical and experimental
studies show that unique properties of graphene, such as linear dispersion
relation near the Dirac point, high electron Fermi velocity, and tunable Fermi
level lead to a strong nonlinear response in 2D graphene structures and suggest
it is a very promising candidate for THz applications \cite{wright09,coherent10,sarma11,SHG,
maeng2012gate,4wavegr12,gullans2013single,kumar2013third,review2014,noepgr15,cheng2,mik16}.
This work opens up the possibility of graphene based devices for \si{THz} optoelectronic and photonic applications \cite{grreview,review2014}.\par

Following the first experimental of the study of absorption in the
ellipsometric  spectrum of graphene \cite{kravets10},
the circular AC Hall effect \cite{oka09,CPL1,jiang11}, chiral edge currents
\cite{cpledge10,CPL2}, helicity-dependent photovoltaic Hall effect \cite{oka09,review2014}  and electronic chirality and berry phases \cite{CPL3} were observed
by using circularly-polarized excitation fields. Higher-order
harmonic generation \cite{SHG,sorngaard13} in 2D
graphene has also been theoretically investigated, showing strong higher-order
harmonics exists in graphene with an applied circularly-polarized harmonic
electric field. This work shows that elliptically- or circularly-polarized light may be
used to probe the unique nature of graphene near the Dirac
points, including effects such as harmonic generation, frequency mixing, optical
rectification, linear and circular photogalvanic effect, photon drag,
photoconductivity, coherently controlled ballistic charge currents, pseudospin,
chirality, and symmetry breaking \cite{SHG,kravets10,oka09,CPL1,jiang11,CPL2,CPL3,review2014}. 

In general, graphene nanoribbons (GNRs) have two types of edges: armchair
edges (acGNR) and zigzag edges (zzGNR). Due to the
the geometry and boundary conditions \cite{fertig1,fertig2,and1}, these two
types of GNRs show distinct electronic characteristics in
the low energy regime. The linear and nonlinear response of GNRs due to a
linearly-polarized electric field were studied in
\cite{liao08optical,duan2010infrared,gnrselectionrule,floquetspectrum,gnrcpl2,arxiv1,arxiv2}.
The non-perturbative DC conductance due to an applied circularly-polarized field
in zzGNR and acGNR \cite{floquetspectrum,gnrcpl2} was also investigated.
However, there has been no investigation of the nonlinear response
in metallic acGNR for an elliptically polarized
applied electric field.

In this Letter, we describe new results on the nonlinear response of intrinsic and extrinsic
metallic acGNR
(mGNR) excited by a normally-incident, \textit{elliptically-polarized} THz electric
field. GNR are metallic in the $\mathbf{k}\cdot\mathbf{p}$ approximation
when the longitudinal direction ($\hat{y}$) of the nanoribbon is
parallel to the armchair edge and the nanoribbon atomic width is $N =
(3M-1)$ with $M$ odd.
In this case the lowest sub-bands are linear in $k_y$, and for sufficiently
narrow mGNRs $(L_x < \, \sim \SI{20}{nm})$, the higher sub-bands are far enough away that their contributions to
the THz nonlinear conductance may be neglected.
\textit{Most significantly}, we show that at room temperature, the
third-order nonlinear conductance at $3\omega$ is enhanced by $2-3$ orders of
magnitude using a circularly-polarized (CP) THz field over the same conductance
when the excitation field is linearly-polarized (LP)\cite{arxiv1}. We also show
that the third-order conductances at $\omega$ and $3\omega$
exhibit odd symmetry in the polarization state, 
resulting in current densities of
opposite sign for opposite-handed elliptical polarizations. Finally, we analyze the
Fermi level and temperature dependence of these nonlinearities and show that by
varying the polarization state of the excitation field it is possible to tune the
nonlinearities in both sign and amplitude. This novel behavior suggests a
variety of applications in optical modulation, polarization switching, 
and harmonic generation over the THz region of the optical spectrum.

Our $\mathbf{k}\cdot\mathbf{p}$ model
employs Fourier analysis to solve the nonlinear Dirac equation using
time-dependent perturbation theory in order to study the \textit{polarization-dependent}
nonlinear response in the THz regime. The model is an extension of one first
applied by us in the context of graphene nanoribbons for the study of nonlinear effects
induced by LP THz fields \cite{arxiv1,arxiv2}. However, we emphasize that the
work presented below contains new physics as a result of the coupling between
the polarization state of the incident field and the chiral mGNR
wavefunctions, and is not a simple superposition of the previous description of
LP THz excitations (although the previous
results do exist as special cases of the current work).

In the following, we analyze nonlinear harmonic generation at THz frequencies
induced by an elliptically-polarized beam normally-incident on a mGNR.
The polarization ellipse is characterized by major and minor axes that coincide
with the longitudinal and transverse axes of the nanoribbon.
This polarization state can be achieved experimentally
by passing a $\hat{y}$-polarized beam through a cascade of a half-wave
plate oriented with its fast axis at an angle $\phi/2$ with
respect to the polarization axis of the
incident THz beam, and then through a quarter-wave plate oriented with its
principle axes parallel to the longitudinal $(\hat{y})$ and transverse
$(\hat{x})$ axes of the mGNR. The corresponding electric field may be expressed as 
 $\mathbf{E}=\left[i\hat{x}E_x+\hat{y}E_y\right]\exp[-i\omega t]=E_0\left[i \hat{x} \sin(\phi)+\hat{y} \cos(\phi) \right]\exp[-i\omega t]$.

We begin by writing the polarization state of an
elliptically-polarized beam with
principal axes parallel to $\hat{x}$ and $\hat{y}$.
In the Coulomb gauge for a source-free region of constant scalar potential
$(\nabla \varphi = 0)$,
the time-harmonic electric field turns on adiabatically at $t_0 = -\infty$ and the magnetic vector potential is $\mathbf{A}=\mathbf{E}/(i\omega)$. After making the substitution,
$\mathbf{k(k')}\to\mathbf{k(k')}+q\mathbf{A}/\hbar$, we obtain a
time and polarization-state dependent Hamiltonian $H$ near the Dirac points of the mGNR
\cite{arxiv1,arxiv2}. The Fourier expansion of the resulting perturbation wave
function is written:
\begin{equation*}
\label{eq:psi}
\psi(\mathbf{r}, t;m)=\sum_{l=0}^{\infty} \psi_0(m,l) \exp\left[i 2\pi m y/L_y\right] \exp[- i
\omega l t] \exp[-i \epsilon t / \hbar]
\end{equation*}
where $m$ is the quantum number of $k_y$, $l$ is the harmonic
order of the electric field, and
$\psi_0(m,l)$ is a spinor of order $(m,l)$.
Following Ref. \cite{review2014,wright09,cheng2,mik16,arxiv1}, we obtain the third order current density in the perturbation limit, with $\omega_y=v_F 2\pi m/L_y$:
\begin{subequations}
\begin{align}\label{eq:j}
J_{\nu}^{(3)}(\phi,t)=&\sum_{\alpha\beta\gamma}\left[\exp\left[-i\omega t\right]
\sigma_{\nu\alpha\beta\gamma}^{(3)}(\omega,-\omega,\omega) E_{\alpha}E_{\beta}^{*}E_{\gamma}\right. \nonumber \\
&+\exp\left[-i\omega t\right] \sigma_{\nu\alpha\beta\gamma}^{(3)}(\omega,\omega,-\omega) E_{\alpha}E_{\beta}E_{\gamma}^{*}\\
&+\exp\left[-i3\omega t\right]
\left.\sigma_{\nu\alpha\beta\gamma}^{(3)}(\omega,\omega,\omega) E_{\alpha}E_{\beta}E_{\gamma}\right]+c.c. \nonumber\\
=&\mathrm{e\, g_s \,g_v}\sum_{m}\left[\psi_0(m,1)^\dagger \frac{\partial
H}{\hbar \partial  k_\nu}\psi_0(m,2)\exp[-i\omega t]\right. \\
&\left.+\psi_0(m,0)^\dagger \frac{\partial H}{\hbar \partial  k_\nu} \psi_0(m,3)\exp[-i3\omega t]\right]N(\omega_y)+c.c. \nonumber\\
=&\left[g_{\nu}^{(3)}(\omega,\phi)\exp[-i \omega t]\right. \\
&\left.+g_{\nu}^{(3)}(3\omega,\phi)\exp[-i 3 \omega t]\right]E_0+c.c.\nonumber
\end{align}
\end{subequations}
with the thermal factor defined as:
\begin{equation}
N (\omega_y)=\frac{\sinh \left( \frac{\hbar |\omega_y|}{k_B T} \right)}{\cosh \left( \frac{E_F}{k_B T} \right)+\cosh \left( \frac{\hbar |\omega_y|}{k_B T} \right)} 
\end{equation}
with $\nu = x, y$ indicating the induced optical current component in the $\hat{\nu}$ direction. The longitudinal Kerr conductance $g_{\nu}^{(3)}(\omega,\phi)$ and third-harmonic conductance $g_{\nu}^{(3)}(
3\omega,\phi)$ for infinitely-long mGNR:
\begin{subequations}
\begin{align}
\label{eq:gyw}
g_{y}^{(3)}(\omega,\phi) =g_0  &\left[f(\phi,-2,-1)N\left(\frac{\omega}{2}\right)+f(\phi,-1,-\frac{3}{2}) N(\omega)\right]\\
\label{eq:gy3w}
g_{y}^{(3)}(3\omega,\phi,\lambda)=g_0  &\left[ f(\phi,\frac{1}{2},-\frac{1}{24})N\left(\frac{\omega}{2}\right) -f(\phi,1,\frac{5}{6})N(\omega) \right.\nonumber\\
&\left.+f(\phi,\frac{1}{2},\frac{7}{8})N\left(\frac{3\omega}{2}\right)\right]
\end{align}\label{eq:gy}
\end{subequations}
and the transverse third-order conductances:
\begin{subequations}
\begin{align}
\label{eq:gxw}
g_{x}^{(3)}(\omega,\phi) =g_0 &\left[f(\phi,1,-\frac{1}{2}) N(\omega)\right]\\
\label{eq:gx3w}
g_{x}^{(3)}(3\omega,\phi,\lambda)=g_0 &\left[-f(\phi,\frac{1}{2},-\frac{5}{24})N\left(\frac{\omega}{2}\right) +f(\phi,1,\frac{1}{6})N(\omega) \right.\nonumber\\
&\left.-f(\phi,\frac{1}{2},\frac{7}{8})N\left(\frac{3\omega}{2}\right)\right]
\end{align}\label{eq:gx}
\end{subequations}
where $f(\phi,a,b)=\eta\eta_x\cos(\phi)\left [a \cos^2(\phi)+2b\sin^2(\phi)\right]$
, $g_0=e^2/(4\hbar^2)$, Fermi level $E_F$,
$\eta_{x}=\left(g_s g_v v_F\right)/\left(\omega L_x\right)$, and
$\eta=\left(e^2E_0^2v_F^2\right)/(\hbar^2\omega^4)$.
It is worth noting that for a circularly-polarized excitation field,
a symmetry-breaking occurs in 2D SLG that allows second-harmonic generation to occur
\cite{review2014,SHG,cheng2}.
We will discuss how this symmetry-breaking affects second-harmonic generation in mGNR in a future paper.\par
In \cref{eq:gyw,eq:gxw}, $E_0 f(\phi,a,b)$ may be split into two terms: 
\begin{equation*}
E_0 f(\phi,a,b) = f_A(\phi,a) |E_y|^2 E_y +
f_B(\phi,b) P_{circ} E_x,
\end{equation*}
where the radiation helicity $P_{circ} = i \left [ E_y
(i E_x)^* - i E_x E_y^* \right ] = E_0^2 \sin(2 \phi)$ \cite{cpledge10}.
$f_A(\phi,a)$ defines the conductivity tensor element $\sigma_{\nu yyy}^{(3)}$
and $f_B(\phi,b)$ defines the sum of conductivity tensor elements
$\sum_{yxx} \sigma_{\nu yxx}^{(3)}$, where $\sum_{yxx}$ indicates the sum
over the rotation of indices $\nu yxx$, $\nu xyx$, and $\nu xxy$. We summarize
the tensor elements obtained from \cref{eq:gyw,eq:gxw} in \cref{table:w}.\clearpage
\begin{table}[htpb]
\centering
\caption{Kerr conductivity tensor elements}
\begin{tabular}{ c c }
\hline\hline    
\vtop{\hbox{single-photon $(-\omega/2 \to \omega/2)$}\hbox{conductivity
element(s)}} & \vtop{\hbox{value}\hbox{($\times g_0 \eta \eta_x/E_0^2$)}}\\
\hline
$\sigma_{yyyy}^{(3)}(\omega,-\omega,\omega)$ & $-2$\\
$\sigma_{yxxy}^{(3)}(\omega,-\omega,\omega)+\sigma_{yyxx}^{(3)}(\omega,-\omega,\omega)$ & $-2$\\
$\sigma_{yxyx}^{(3)}(\omega,-\omega,\omega)$ & $0$\\
\hline
$\sigma_{xyyy}^{(3)}(\omega,-\omega,\omega)$ & $0$\\ $\sigma_{xxxy}^{(3)}(\omega,-\omega,\omega)+\sigma_{xyxx}^{(3)}(\omega,-\omega,\omega)$ & $+1$\\
$\sigma_{xxyx}^{(3)}(\omega,-\omega,\omega)$ & $-1$\\
\hline\hline    
\vtop{\hbox{two-photon $(-\omega \to \omega)$}\hbox{conductivity element(s)}} &
\vtop{\hbox{value}\hbox{($\times g_0 \eta \eta_x/E_0^2$)}}\\
\hline
$\sigma_{yyyy}^{(3)}(\omega,\omega,-\omega)$ & $-1$\\
$\sigma_{yxyx}^{(3)}(\omega,\omega,-\omega)+\sigma_{yyxx}^{(3)}(\omega,\omega,-\omega)$ & $-2$\\
$\sigma_{yxxy}^{(3)}(\omega,\omega,-\omega)$ & $-1$\\
\hline
$\sigma_{xyyy}^{(3)}(\omega,\omega,-\omega)$ & $+1$\\
$\sigma_{xxyx}^{(3)}(\omega,\omega,-\omega)+\sigma_{xyxx}^{(3)}(\omega,\omega,-\omega)$ & $-2$\\
$\sigma_{xxxy}^{(3)}(\omega,\omega,-\omega)$ & $+1$\\
\hline\hline
\end{tabular}\label{table:w}
\end{table}
For the third-harmonic conductances, the splitting also holds for $E_0
f(\phi,a,b) = f_A(\phi,a) E_y^3 + f_B(\phi,b) E_x^2 E_y/2$. We note that
$f_A(\phi,a)$ and $f_B(\phi,b)$ are related to the conductivity tensor elements
in the same way as above.
We summarize the contributions to the third-harmonic conductivity tensor
elements from the single-photon, two-photon, and three-photon transitions
obtained from
\cref{eq:gy3w,eq:gx3w} in \cref{table:3w}.
\begin{table}[htpb]
\centering
\caption{Third-harmonic conductivity tensor contributions}
\begin{tabular}{ c c }
\hline\hline    
\vtop{\hbox{single-photon ($-\omega/2 \to \omega/2$)}\hbox{conductivity component}} &
\vtop{\hbox{value}\hbox{($\times g_0 \eta \eta_x/E_0^2$)}}\\
\hline
$\sigma_{yyyy}^{(3^{\prime})}$ & $+1/2$\\
$\sigma_{yxyx}^{(3^{\prime})}+
\sigma_{yyxx}^{(3^{\prime})}+
\sigma_{yxxy}^{(3^{\prime})}$ & $-1/12$\\
\hline
$\sigma_{xyyy}^{(3^{\prime})}$ & $-1/2$\\
$\sigma_{xxyx}^{(3^{\prime})}+
\sigma_{xyxx}^{(3^{\prime})}+
\sigma_{xxxy}^{(3^{\prime})}$ & $+5/12$\\
\hline\hline    
\vtop{\hbox{two-photon $(-\omega \to \omega)$}\hbox{conductivity component}} &
\vtop{\hbox{value}\hbox{($\times g_0 \eta \eta_x/E_0^2$)}}\\
\hline
$\sigma_{yyyy}^{(3^{\prime})}$ & $-1$\\
$\sigma_{yxyx}^{(3^{\prime})}+
\sigma_{yyxx}^{(3^{\prime})}+
\sigma_{yxxy}^{(3^{\prime})}$ & $-5/3$\\
\hline
$\sigma_{xyyy}^{(3^{\prime})}$ & $+1$\\
$\sigma_{xxyx}^{(3^{\prime})}+
\sigma_{xyxx}^{(3^{\prime})}+
\sigma_{xxxy}^{(3^{\prime})}$ & $+1/3$\\
\hline\hline    
\vtop{\hbox{three-photon $(-3\omega/2 \to 3\omega/2)$}\hbox{conductivity
component}} & \vtop{\hbox{value}\hbox{($\times g_0 \eta \eta_x/E_0^2$)}}\\
\hline
$\sigma_{yyyy}^{(3^{\prime})}$ & $+1/2$\\
$\sigma_{yxyx}^{(3^{\prime})}+
\sigma_{yyxx}^{(3^{\prime})}+
\sigma_{yxxy}^{(3^{\prime})}$ & $+7/4$\\
\hline
$\sigma_{xyyy}^{(3^{\prime})}$ & $-1/2$\\
$\sigma_{xxyx}^{(3^{\prime})}+
\sigma_{xyxx}^{(3^{\prime})}+
\sigma_{xxxy}^{(3^{\prime})}$ & $-7/4$\\
\hline\hline
\multicolumn{2}{l}{
Here $\sigma_{\nu \alpha \beta \gamma}^{(3^{\prime})}$
denotes $\sigma_{\nu \alpha \beta \gamma}^{(3)}(\omega,\omega,\omega)$} \\
\end{tabular}\label{table:3w}
\end{table}\par
\begin{figure}[htpb]
\centering
\subfloat{\label{fig:1a}\includegraphics[width=0.5\linewidth]{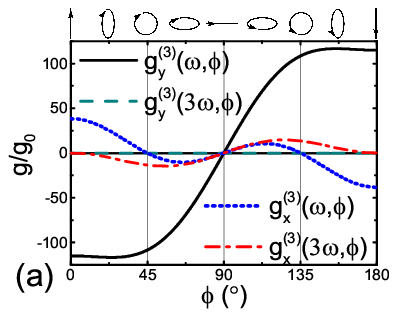}}
\hfil
\subfloat{\label{fig:1b}\includegraphics[width=0.5\linewidth]{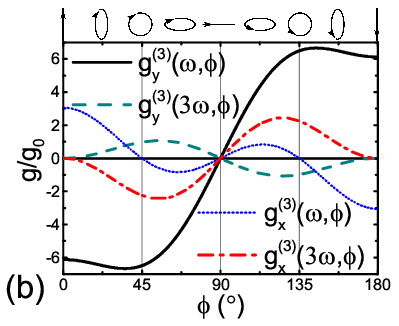}}
\caption{\label{fig:1}Kerr and third-harmonic conductances of mGNR20 as a function of the incident
THz electric field polarization state (see top inset) for: (a) $T=\SI{0}{K}$, and (b)
$T=\SI{300}{K}$. $f=\SI{1}{THz}$, $E_0 = \SI{10}{kV/m}$.}
\end{figure}
\begin{figure}[htpb]
\centering
\includegraphics[width=0.8\linewidth]{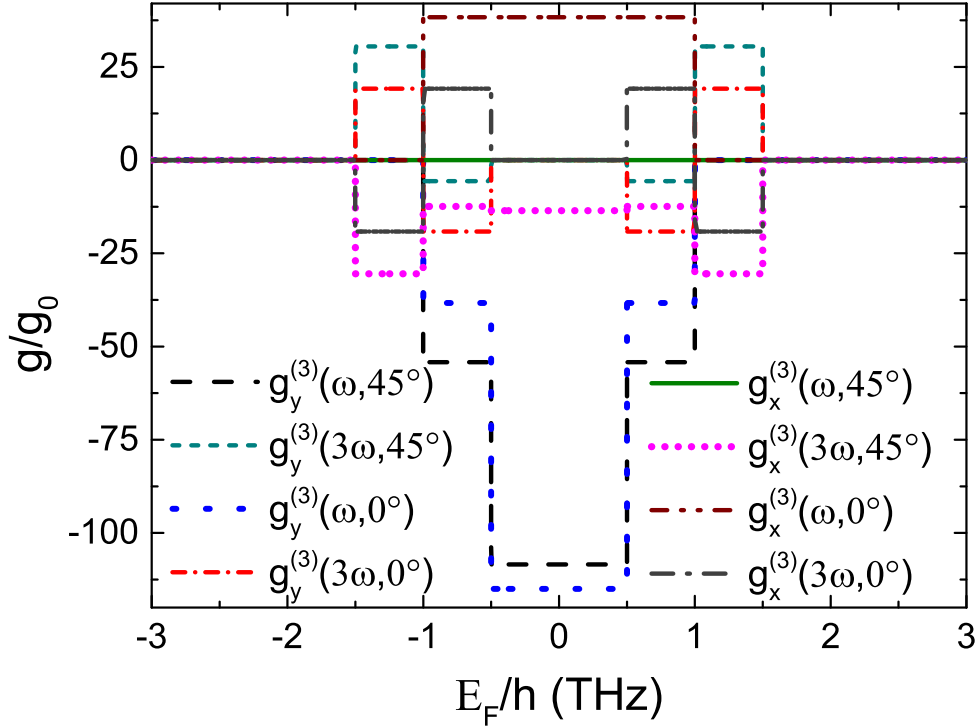}
\caption{\label{fig:2}Longitudinal and transverse components of the Kerr and third-harmonic
conductances for extrinsic mGNR20 as a function of the Fermi level $E_F$ and
incident field polarization state. For this plot $f=\SI{1}{THz}$, $E_0 =\SI{10}{kV/m}$,
and $T=\SI{0}{K}$.}
\end{figure}

To simplify the discussion, in the following we present results for mGNR20, the
metallic acGNR $N=20$ atoms wide, for an applied field strength
$E_0=\SI{10}{kV/m}$.
From \cref{eq:gy,eq:gx}, we see that illumination of an unbiased, infinitely-long
mGNR by a \si{THz} harmonic electric field results in a nonlinear
response that is strongly dependent on the polarization state of
the applied field. \cref{fig:1} illustrates the polarization
dependence of the longitudinal and transverse Kerr and third-harmonic nonlinear
conductances at $T = \SI{0}{K}$ and \SI{300}{K} for intrinsic mGNR20.
\\
For CW ($\sigma_+$) and CCW ($\sigma_-$) CP, \textit{i.e},
$\phi=\ang{45}$ and \ang{135}, the third-order conductances are
antisymmetric. 
The shape of the conductance is
a superposition of the contribution from the $E_y$ and $E_x$ components. 
The overall dependence of the third order current component $J_{\nu}^{(3)}(\omega_0)$ is of the form \cite{wright09,arxiv1},
\begin{equation}\label{eq:gphi}
J_{\nu}^{(3)}(\omega_0)=g_{\nu}^{(3)}(\omega_0,\phi)E_0 =g_{\nu, A}(\omega_0) E_y+g_{\nu, B}(\omega_0)E_x
\end{equation}
The first term, $g_{\nu, A}\propto e^2 E_y^2/(\hbar^2 \omega^4)$, and agrees with \cite{review2014,wright09,cheng2,mik16}.
The second term, $g_{\nu, B} \propto P_{circ}$ (see \textit{e.g.}
\cite{cpledge10,CPL2}). Due to the current operator used, there is no analog of
left or right handedness for the carriers in mGNR \cite{klein}. The direction of
the {\it optically induced third order current} results from the interference
between the local current density excited by the elliptical polarization of the field, or the radiation helicity $P_{circ}$, and the isospin of the carriers \cite{review2014,cpledge10,CPL3,klein}. Finally, for the LP cases $\hat{y}$ and $\hat{x}$, $(\phi=\ang{0}$ and $\phi=\ang{90})$ respectively, contribution from the second term vanishes \cite{CPL3} and we recover the results reported previously \cite{arxiv1}.
In \cref{fig:2} we compare the longitudinal and transverse Kerr
and third-harmonic conductances
as a function of the Fermi level $E_F$ at $T=\SI{0}{K}$. For $E_F$ well below
the optical phonon energy ($\sim$ \SI{200}{meV}), and a 1 THz excitation
frequency for a
LP field in the $\hat{y}$ direction ($\phi=\ang{0}$) and
$\sigma_+$ CP ($\phi=\ang{45}$) we see thresholding behavior of the nonlinear
conductances for the direct interband transition at low temperature \cite{arxiv1}. The three critical frequencies for $E_F/h$: \SI{0.5}{THz}, \SI{1}{THz} and \SI{1.5}{THz} correspond to turning off the
thermal distribution at $\omega/2$, $\omega$ and $3\omega/2$ \cite{arxiv1}.
These frequencies are independent of the polarization states $\phi$, and are only functions of $g_{\nu,A}(\omega_0)$ and$g_{\nu,B}(\omega_0)$ in \cref{eq:gphi}. \par
In the interest of brevity, we note briefly that
the overall temperature dependence of all non-zero extrinsic conductances show that the
nonlinearity persists even up to room temperature.  Further,
the curves asymptotically
approach the intrinsic mGNR conductance for a given polarization state as the
temperature increases. It is also interesting to note that the transverse Kerr
conductances are identically zero for CP, independent of the Fermi
level and temperature, in qualitative agreement with \cite{gnrcpl2}.
This may be understood as the cancellation of the contribution from $E_y$ and $P_{circ}$
in \cref{eq:gxw} for CP.\par

\begin{figure}[htpb]
\centering
\subfloat{\label{fig:3a}\includegraphics[width=0.45\linewidth]{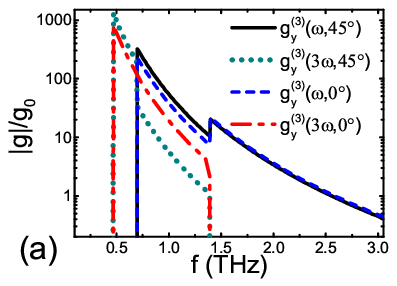}}
\hfil
\subfloat{\label{fig:3b}\includegraphics[width=0.45\linewidth]{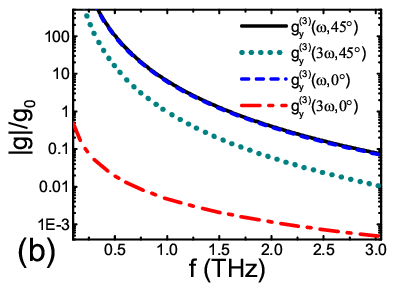}}
\vfil
\subfloat{\label{fig:3c}\includegraphics[width=0.45\linewidth]{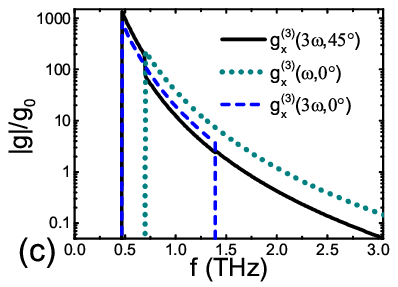}}
\hfil
\subfloat{\label{fig:3d}\includegraphics[width=0.45\linewidth]{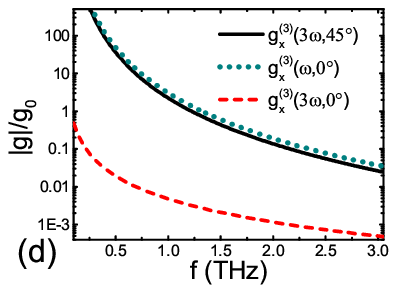}}
\caption{\label{fig:3}Excitation frequency ($2 \pi f = \omega$) dependence of the components of nonlinear
Kerr and third-harmonic conductances for extrinsic mGNR20 ($E_F/h =
\SI{0.7}{THz}$): a) longitudinal conductance at $T=\SI{0}{K}$, b) longitudinal
conductance at $T=\SI{300}{K}$, c) transverse conductance at $T=\SI{0}{K}$, and
d) transverse conductance at $T=\SI{300}{K}$. For all plots, $E_0 = \SI{10}{kV/m}$. Note that for all temperatures, the CP transverse Kerr conductance is identically zero.}
\end{figure}

\cref{fig:3} illustrates the excitation-frequency $(2 \pi f = \omega)$
dependence of the Kerr and third-harmonic conductances for both $\hat{y}$ LP
$(\phi=\ang{0})$ and $\sigma_+$ CP $(\phi=\ang{45})$ excitation fields in
extrinsic mGNR20
$(E_F/h=\SI{0.7}{THz})$. At $T=\SI{0}{K}$ (\cref{fig:3a,fig:3c}) both CP Kerr
conductance components behave in a manner qualitatively similar to the LP result for the
longitudinal component of the conductance while the transverse component of the
LP conductance is zero. However
the CP third-harmonic conductances behave quite differently from their LP
counterparts. Whereas the low-temperature LP conductance components
are bandlimited (nonzero over $2 E_F/3 h
< f < 2 E_F/h$), the transverse CP conductances \textit{persist to significantly higher
frequencies}, reducing to $|g|/g_0 = 0.1$ at approximately $\SI{2.7}{THz}$. At $T =
\SI{300}{K}$ (\cref{fig:3b,fig:3d}), another behavior is noted: while the CP and
longitudinal component of the LP Kerr conductance
follow a similar decay envelope with increasing excitation
frequency,
both CP third-harmonic conductance components are \textit{enhanced by nearly three
orders of magnitude} over their LP counterparts at $f = \SI{1}{THz}$, and this
enhancement persists to higher frequencies and is
still nearly two orders of magnitude at $f = \SI{3}{THz}$. The
enhancement for the transverse component of the CP third-harmonic conductance is observed to be
slightly stronger than that for the longitudinal component of the CP third-harmonic conductance.
\par
In conclusion, we report calculations describing the third-order nonlinear
response of both intrinsic and extrinsic mGNR to an elliptically-polarized THz
electric field. We show that the resulting Kerr conductances for extrinsic mGNR
persist to significantly higher excitation frequencies at low temperature, and
at room temperature, the CP third-harmonic conductances are enhanced by 2-3
orders of magnitude over their counterparts excited with only LP. Further, we
describe the Fermi-level and temperature dependence of these nonlinearities. The
enhancement in spectral range and magnitude for these nonlinearities suggests
that they may exhibit wide applicability in THz devices excited with
elliptically-polarized THz electric fields.

The recent synthesis of ultrathin mGNR with widths $L_x < \SI{10}{nm}$
\cite{kimouche2015ultra,jacobberger2015direct}, coupled with the proposed
experimental setup described in this letter, suggest that experimental
measurement of the \si{THz} nonlinear response in thin mGNR
should be possible at relatively low excitation field strengths. Notably, the
enhancement of the third-order third-harmonic nonlinearity with small changes in
Fermi level and applied CP excitation field at room temperature indicates that mGNR may provide the basis for developing a sensitive graphene-based detector,
broadband modulator, or source over a wide range of temperatures.

\bibliographystyle{apsrev4-1}
%

\end{document}